\documentclass[aps,twocolumn,superscriptaddress,showpacs,prb]{revtex4}
\usepackage{epsfig}

\begin{document}

\title{Liquid-gas separation in colloidal electrolytes}

\author{Jos\'e B. Caballero}
\affiliation{Group of Complex Fluids Physics, Department of Applied Physics, University of Almeria, 04120 Almeria, Spain}
\author{Antonio M. Puertas}
\email[Corresponding author: ]{apuertas@ual.es}
\affiliation{Group of Complex Fluids Physics, Department of Applied Physics, University of Almeria, 04120 Almeria, Spain}
\author{Antonio Fern\'andez-Barbero}
\affiliation{Group of Complex Fluids Physics, Department of Applied Physics, University of Almeria, 04120 Almeria, Spain}
\author{F. Javier de las Nieves}
\affiliation{Group of Complex Fluids Physics, Department of Applied Physics, University of Almeria, 04120 Almeria, Spain}
\date{\today}
\author{J.M. Romero-Enrique}
\affiliation{Depto. de F\'{\i}sica At\'omica, Molecular y Nuclear, Area de F\'{\i}sica Te\'orica, Universidad de Sevilla, Aptdo. 1065, 41080 Sevilla, Spain}
\author{L.F. Rull}
\affiliation{Depto. de F\'{\i}sica At\'omica, Molecular y Nuclear, Area de F\'{\i}sica Te\'orica, Universidad de Sevilla, Aptdo. 1065, 41080 Sevilla, Spain}

\begin{abstract}

The liquid-gas transition of an electroneutral mixture of oppositely charged colloids, studied by Monte Carlo simulations, is found in the low temperature -- low density region. The critical temperature shows a non-monotonous behavior as a function of the interaction range, $\kappa^{-1}$, with a maximum at $\kappa \sigma \approx 10$, implying an island of coexistence in the $\kappa$-$\rho$ plane.
The system is arranged in such a way that each particle is surrounded by shells of particles with alternating charge. In contrast with the electrolyte primitive model, both neutral and charged clusters are obtained in the vapor phase.

\end{abstract}


\maketitle

\section{Introduction}

Electrostatic correlations play an important role in model, biological and applied systems \cite{levin}, being their comprehension a huge challenge for the liquid state researchers. While progress has been made in understanding the influence of those correlations in simpler ionic systems, only recently the importance of macromolecular correlations has been acknowledged. Furthermore, for charged colloidal systems, the interaction depends not only on the charge density and distribution, but also on the solvent properties; the interactions are thus tunable by acting on it (added salt, temperature...). This is the reason why phase diagrams are, in general, richer in colloidal systems \cite{rull95}.

The Restricted Primitive Model (RPM) for ionic fluids is the simplest mixture showing strong effects due to charge correlations \cite{mcquarrie76}. This model undergoes a liquid-gas transition at low temperature and density due to the strength of the correlations \cite{levin, romero02} with Ising-like behavior \cite{kim03}, despite the long range of the interactions. This model has been also extended to tackle size or charge asymmetries\cite{jm00,yan02,rescic01,hynninen05}, as an approach to charged colloids. The phase diagram of dipolar systems has been also studied, where a vapor-liquid coexistence island is found as a function of 
the dipolar strength \cite{mcgrother96}.

In this work, we study the liquid-gas transition in the colloidal analogue of the RPM, using a mixture of oppositely charged colloids by means of computer simulations. We will focus on the case in which the concentrations of both colloidal
species is the same, even though electroneutrality can be obeyed when the solution is not equimolar
due to the electrolyte in the medium. Nevertheless this is the case where the correlation effects between unlike colloidal particles will be more significant. To date, the experimental works have focused on the crystal phases \cite{blaaderen05,bartlett05}, where superlattice crystals were reported, and recent simulations concentrated in the clustering of the particles \cite{ryden05}. In a previous work, the liquid-gas transition was indeed found for this system in the low density -- low temperature region\cite{caballero04}, and we concentrate here in the effect of the range of the electrical interaction among particles, experimentally controlled by the electrolyte concentration in the medium. 

We find an unexpected non-monotonous behavior for the critical temperature with the range of interaction, with a maximum at $\kappa \sigma \approx 10$, thus showing a re-entrance phenomenon (fluid -- phase separation -- fluid) by increasing the salt concentration. This is a collective effect since a third term in the virial expansion of the pressure is necessary to describe the non-monotonous critical temperature curve. This fact is then rationalized from the strong correlation between oppositely charged particles, leading in both phases to particles surrounded by shells of alternating charges and clusters growing in the dilute phase. Under this configuration, repulsive and attractive interactions are screened in a different way, provoking  a maximum in the critical temperature. Finally, we show that an increase of the amount of clusters is not a signature of the proximity to the phase transition. That is actually driven by the energy gain in the dense phase to overcome the entropy lost in forming a liquid. 

The paper is organized as follows: Section II describes the details of the model and the simulation techniques. Section III is devoted to present and discuss the results of this work. Finally, the main conclusions can be found in Section IV.

\section{Model and Simulations}

We simulate a 1:1 binary mixture of $N$ spherical colloidal particles, $N/2$ bearing a surface potential $+\phi$ and $N/2$ with $-\phi$. The interaction between colloids is modeled by the effective DLVO electrostatic interaction \cite{vo48}, plus a hard-core repulsion,

\begin{equation}
V(r_{ij})\:=\:V_{HS}(r_{ij})\,+\,\phi_i\phi_j \exp\left\{-\kappa(r_{ij}-\sigma)\right\}
\end{equation}

\noindent where $V_{HS}$ is the hard-sphere potential for a particle of diameter $\sigma$, $\phi_i$ and $\phi_j$ are the surface potentials of the interacting particles (in appropriate units) and $\kappa$ is the inverse Debye length. This is obviously an approximation to the real problem, where the ions are not simulated, but an effective interaction between the colloids is used. This method, however, allows to study phase separation with large systems, as compared to those where the ions are explicitly included \cite{ryden05}. We shall use in this work reduced units: $\sigma=1$, $U^*=U/\phi^2$, $T^*=k_BT/\phi^2$ and the density $\rho^*=N \sigma^3/V$, where $N$ is the number of particles, and $V$ the volume of the system.

Monte Carlo simulations have been used to compute the gas-liquid coexistence curve. First, Gibbs Ensemble Monte Carlo (GEMC) simulations with $N=432$ particles were run for different $\kappa\sigma$ values. Equilibration of the system takes very long times ($1-5 \cdot 10^6$ cycles); however, contrary to the RPM, multiparticle moves do not speed up the the equilibration rate due to the high density of the liquid phase. Production runs comprised at least $10^6$ cycles, taken at equilibrium.

Grand Canonical Monte Carlo (GCMC) simulations aided with reweighting techniques\cite{ferrenberg,ferrenberg2} were used to verify the results from GEMC. Owing to the long range of the correlations, the coexistence curve depends strongly on the system size in the neighborhood of the critical point. This fact can be used to locate the critical point using the mixed-field finite size scaling analysis developed by Bruce and Wilding \cite{bruce92}. This method is based on the asymmetry of the density distribution, reflecting the absence of particle-hole symmetry in off-lattice models. Thus, the order parameter of the transition is not the density, but a mixture between the density and the energy: $M \sim \rho+su$ (where $s$ is a system-dependent field mixing parameter). Precisely at criticality, the distribution of $M$ in a large enough system, with box size $L$, takes on the universal form: 
\begin{equation}
P_L(M)=a_L P^*\left(a_L\left[M-\langle M \rangle\right]\right)
\end{equation}
where $P^{*}(x)$ is an universal distribution function for each universality class, $<\dots>$ is the average evaluated at critical conditions and $a_{L} \sim L^{\beta/\nu}$. Thus, using GCMC simulations, critical parameters for different box sizes are calculated, and applying the corresponding scaling laws, the critical parameters for an infinite system can be obtained 
\begin{eqnarray}
T_c(L)-T_c(\infty) \sim L^{-(\theta+1)/\nu}\\
\rho_c(L)-\rho_c(\infty) \sim L^{-(1-\alpha)/\nu}
\end{eqnarray}
where $\nu$ and $\alpha$ are the critical exponents associated to the correlation length and heat capacity divergence, respectively, and $\theta$ is the correction-to-scaling exponent. For the 3D Ising universality class, $\alpha\approx 0.11$, $\nu \approx 0.629$ and $\theta\approx 0.54$ \cite{liu}. Some remarks are pertinent at this point. First of all, the Bruce-Wilding method does not identify the universality class of the system but to obtain the critical parameters assuming certain universality class. Since our model presents a short range interaction, 3D Ising criticality is expected. On the other hand, the Bruce-Wilding method is not consistent with the Yang-Yang anomaly. This method can be improved including the pressure as a scaling field, but for Ising-type systems this contribution is usually negligible \cite{fisher04}. Therefore, we follow the Bruce-Wilding method.

To carry out this analysis, we have used GCMC simulations with $L=8$, $9$, $10$ and $12$ system sizes. Runs comprised $25 \cdot 10^6$ configurations for $L=8$ and $L=9$ and $75 \cdot 10^6$ configurations for $L=10$ and $L=12$, configurations are spaced by $2<N>$ attempts to insert or remove a random particle. Statistical errors were taken from three independent simulations runs.

\section{Results and Discussion}

\begin{figure}
\psfig{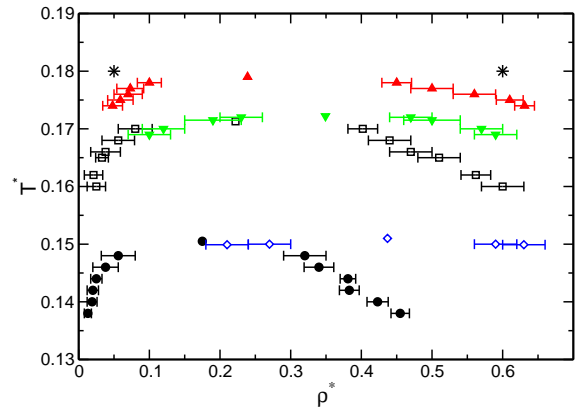}
\caption {\label{phase_diagram} Gas-liquid coexistence points for different values of $\kappa$: $\kappa=3.9$ circles, $\kappa=6$, open squares, $\kappa=10$, upward triangles, $\kappa=15$, downward triangles, and $\kappa=20$, diamonds. The points without error bars mark the critical points, estimated with the law of rectilinear diameters.}
\end{figure}

The liquid-gas transition, studied by means of Gibbs Ensemble Monte Carlo and Grand Canonical simulations, is found in the low-$T$--low-$\rho$ region, where strong correlations between oppositely charged particles emerge, producing a thermodynamic instability \cite{levin,caballero04}. We find liquids and vapors composed of the same number of positive and negative particles on average. The GEMC results for different $\kappa$ are presented in Fig. \ref{phase_diagram}. In contrast to monocomponent attractive systems \cite{poon96}, the critical temperature evolves non-monotonically. At low $\kappa$, demixing occurs at higher temperatures the shorter the interaction range, and only at high $\kappa$, the critical temperature moves to lower values as $\kappa$ increases. Noteworthily, this behavior implies a closed region of phase separation in the constant-$T$, $\kappa$-$\rho$ plane, which is experimentally more accessible than the $T$-$\rho$ one for colloids. A reentrance phenomenon is thus predicted by increasing the salt concentration, from fluid to phase separation and to fluid again, similarly to the dipolar systems \cite{mcgrother96}.

\begin{figure}
\psfig{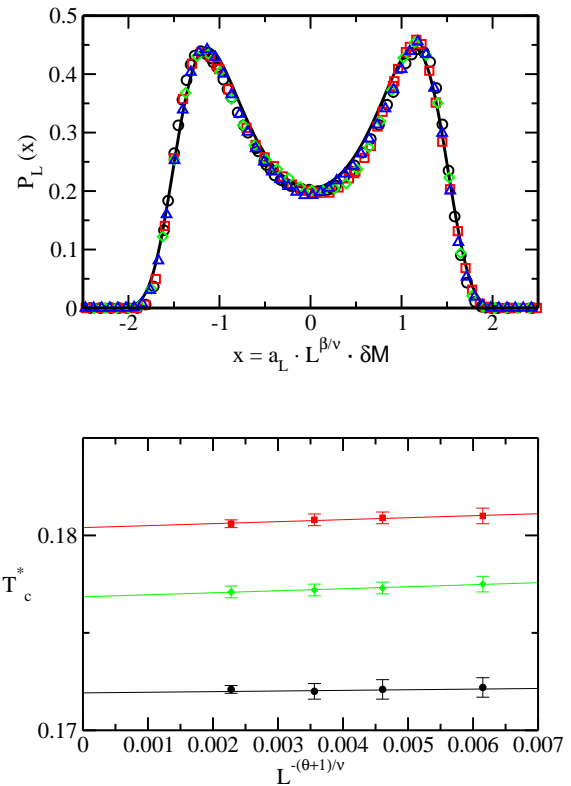}
\caption {\label{plx} Upper panel: order parameter probability distribution $P_{L}(x)$ for $\kappa\sigma=6$ and several box sizes: $L=8$ (black circles), $L=9$ (red squares) $L=10$ (green diamonds) and $L=12$ (blue triangles); solid line for 3D-Ising model. Lower panel: variation of the critical temperature with the box sizes for $\kappa\sigma=6$ (black squares) $\kappa\sigma=10$ (red squares) and $\kappa\sigma=15$ (green diamonds). Lines are lineal fits to the numerical results.}
\end{figure}

To verify the dependence of the critical temperature with the salt concentration, we have also used GCMC simulations what are more precise than the GEMC ones. Fig. \ref{plx} (upper panel) shows $P_L(M)$ matched to the universal 3D Ising distribution for $\kappa\sigma=6$ with different box sizes: $L=8, 9, 10$ and $12$. This matching gives us a critical temperature for each system size, that can be used to obtain $T_c$ in an infinite system as the lineal fit to the suitable scaling law \cite{bruce92} (results presented in the lower panel of Fig. \ref{plx}). Note the small dependence of the critical temperature with the size of the simulation boxes; the critical density also depends slightly on $L$ (not shown). These simulations confirm the behavior of the critical temperature with the salt concentration: $T_c$ increases when $\kappa$ increases at low $\kappa$ ($T_c(\kappa\sigma=6)<T_c(\kappa\sigma=10)$) and decreases at high enough salt concentration ($T_c(\kappa\sigma=15)<T_c(\kappa\sigma=10)$). 

The size dependence of the non universal parameter $a_L \sim L^{\beta/\nu}$ (not shown) permits an estimation of the ratio $\beta/\nu$. The values obtained are similar to those of the Ising model ($\beta/\nu=0.518$) for the three ranges studied: $\beta/\nu=0.523(14), 0.509(23)$ and $0.503(20)$ for $\kappa\sigma=6, 10$ and $15$, respectively. The collapse of the data onto the 3D-Ising curve (solid line) in Fig. \ref{plx}, the agreement of numerical values for the ratios $\beta/\nu$, and the analysis of $a_L$, support the compatibility with 3D-Ising criticality, as expected for short-range potentials.

\begin{figure}
\psfig{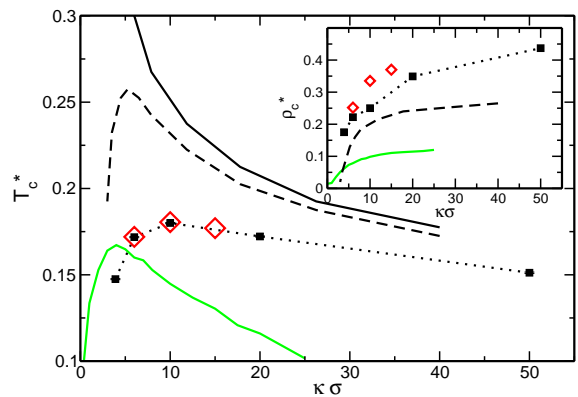}
\caption {\label{parametros} Critical temperature and density (inset) as a function of $\kappa\sigma$: solid squares from GEMC and open red diamonds from GCMC simulations. Black lines are the results using the virial expansion for the pressure up to second (solid line) and third order (broken line) in density. Green lines are results from MSA on a Yukawa mixture \cite{mier-y-teran98}.}
\end{figure}

The critical temperature and density from both GEMC and GCMC simulations are presented in Fig. \ref{parametros} as a function of $\kappa$. The agreement between both simulation results confirms the behavior described above; on the other hand, the critical density, which is better estimated with the GCMC simulations, increases with $\kappa$. The critical parameters have been calculated using the virial expansion of the pressure up to second and third orders (continuous and dashed lines in Fig. \ref{parametros}, respectively): $\beta P(\rho)= \beta P^{CS}_{HS}(\rho)+B_2\rho^2+B_3\rho^3$, where $P^{CS}_{HS}(\rho)$ is the Carnahan-Starling expression for the pressure of hard spheres \cite{hansen90,note3}. The second order expansion produces a liquid-gas transition, driven by the attractions between oppositely charged particles, which behaves monotonically with $\kappa\sigma$ (solid lines). On the other hand, the expansion up to third order correctly reproduces the maximum in $T_c$ vs. $\kappa \sigma$ (dashed lines). Because $B_3$ is needed to reproduce qualitatively the results, the increasing trend of $T_c$ vs. $\kappa\sigma$ at long interaction ranges comes from the interaction between, at least, three particles, mainly the repulsion between two particles with similar charge attracted to a third one. This repulsion is more important for lower $\kappa$, whereas the attraction energy is not greatly increased. The origin of the maximum is, thus, similar to that of the dipolar system \cite{mcgrother96}. These results also agree qualitatively with previous calculations for an attractive-repulsive mixture of Yukawa potentials using an expansion of the internal energy and the equation of state, based on the mean-spherical approximation (MSA) \cite{mier-y-teran98}.

\begin{figure}
\psfig{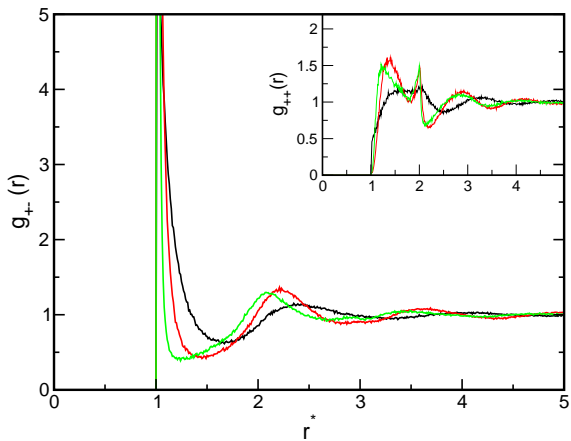}
\caption {\label{denso} Partial pair distribution functions for $T=0.18$ and $\rho=0.6$ with $\kappa=2$ (black curve), $\kappa=6$ (red curve) and $\kappa=15$ (green curve); opposite sign (main figure) and same sign (inset).}
\end{figure}

As in the ionic fluids, the demixing in a dense and a dilute phase is driven by the strong correlations between oppositely charged particles \cite{levin,caballero04}. We have investigated the internal correlations in two supercritical ($T^*=0.18$) states, $\rho^*=0.6$ and $\rho^*=0.05$, marked in Fig. \ref{phase_diagram} as asterisks. Fig. \ref{denso} plots the opposite sign, $g_{+-}(r)$, and same sign (inset), $g_{++}(r)=g_{--}(r)$, contributions to the pair distribution function for the dense state (with $\kappa\sigma=2, 6$ and $15$). The system is composed of layers with alternating sign particles surrounding every particle; first layer of particles of opposite sign, second one of the same sign... This structure extends to distances of several radii, much larger than the interaction range. As observed in the Figure, the peaks shift to shorter distances as $\kappa$ increases. Qualitatively similar features have been predicted using the MSA for the partial structure factors \cite{ruiz-estrada01}.

Second neighbors of one particle comprise the first shell of particles with the same sign, which moves to larger distances as $\kappa$ decreases. This is due to the repulsion between similarly charged particles and correlations through unlike particles. The first neighbor layer, thus, drive the gas liquid transition, whereas the second one impedes it, resulting in a maximum of the critical temperature. A peak at $r=2\sigma$ marks the presence of linear arrangements (three particles long) at all ranges studied. This peak is not caused by the repulsion of the central particle (because the repulsion range is too short), but due to the repulsion of the particles in the second layer, first peak in $g_{++}(r)$. Visual inspection of the system shows that these strings are distributed randomly (thus cannot attributed to crystallites), and are only three particles long (no peak at $r=3\sigma$ in $g(r)$).

\begin{figure}
\psfig{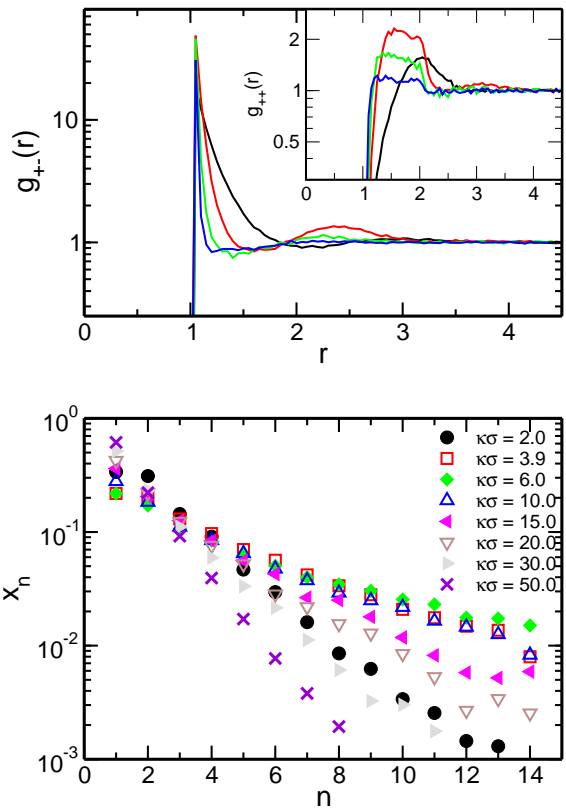}
\caption {\label{diluido} Upper panel: Partial pair distribution functions for $T^*=0.18$ and $\rho=0.05$ with $\kappa=2$ (black curve), $\kappa=6$ (red curve), $\kappa=15$ (green curve) and $\kappa=30$ (blue curve); oppositely sign (main figure) and same sign (inset). Lower panel: Fraction of particles in a cluster with $n$ particles for the same state and different $\kappa$, as labeled. $x_n$ is normalized as: $\sum x_n = 1$.}
\end{figure}

Now we move to the dilute supercritical state $T^*=0.18$ and $\rho^*=0.05$, presented in Fig. \ref{diluido} for different interaction ranges (upper panel). The system is composed of clusters of particles \cite{caballero04}, and the fraction of particles in every cluster, $x_n$, is presented in the lower panel of the figure for different values of $\kappa$. Again the layering of particles inside clusters is observed although only two or three layers are seen due to the finite size of the clusters. The layers move closer as $\kappa$ increases, the effect being more dramatic than in the dense state due to the lower density of the system. Notably, the absence of the peak at $r=2\sigma$ confirms the collective origin of the linear structures found in the dense phase (Fig. \ref{denso}). 

Different from the RPM \cite{bresme95,caillol95}, the distribution of clusters is monotonous, containing both neutral and charged clusters, no matter how close to the transition. However, this distribution shows also a non-monotonic behavior with $\kappa$; for long ranges the number of large clusters increases when $\kappa$ increases, whereas the opposite trend is observed at high $\kappa$, the maximum number of big clusters found for $\kappa\sigma=6.0$ (for not-too-big clusters, the maximum is at $\kappa\sigma=3.9$). These values do not agree with the maximum of $T_c^*$, which indicates that maximal proximity to the transition does not imply maximal tendency to form large clusters in dilute systems. 

\begin{figure}
\psfig{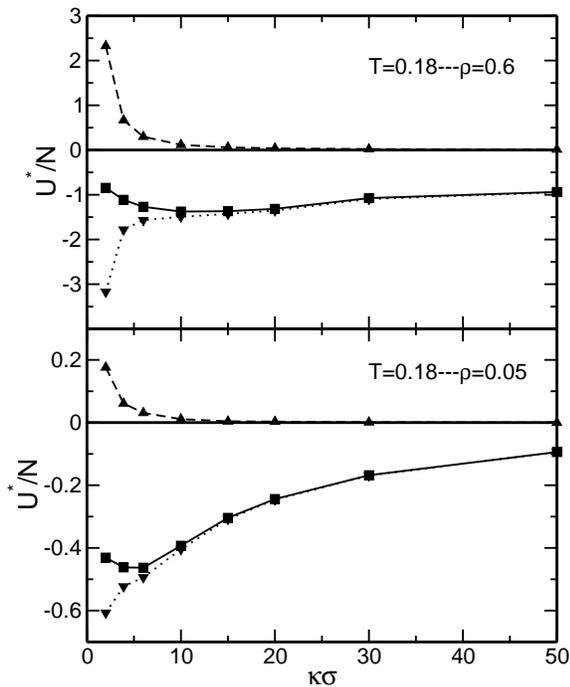}
\caption {\label{ener} Electrostatic energy (squares), and contributions from repulsions (upwards triangles) and attractions (downwards triangles) for two states as labeled.}
\end{figure}

Finally, to complete the understanding of the phase transition and its driving mechanism, Fig. \ref{ener} depicts the internal (electrostatic) energy with its contributions from attractions and repulsions. The same supercritical states are presented ($T^*=0.18$; $\rho^*=0.6$ (upper panel) and $\rho^*=0.05$ (lower panel)). At low $\kappa$, the total energy comprises repulsive and attractive contributions. For larger $\kappa$, however, the energy contains only the attractive contributions due to the different distances between similar sign and opposite sign pairs (see Fig. \ref{denso}). Therefore, the behavior at high $\kappa$ is similar to that of a monocomponent system. For the dilute state, the clustering implies a low number of bonds, as compared to the dense state, resulting in a lower energy (in absolute value). The minimum in the energy can be rationalized using simple small clusters, i.e. trimers and tetramers \cite{caballero05}.

Comparison of the energy curves with the behavior of critical temperature shows that the driving mechanism for liquid-gas separation in this system is the energy gain in forming a dense (liquid) phase, and not the ability to form large clusters in dilute phases (see Fig. \ref{ener}), as proposed for the asymmetric ionic fluids \cite{yan01}. The effect arises from the difference in the number of interacting pairs of similarly charged colloids in the dilute and dense phases. The internal energy calculated using the MSA for a mixture of Yukawa potentials shows similar trends, with minima moving to higher $\kappa$ as the density increases \cite{mier-y-teran98}.

\section{Conclusions}

In conclusion, we have studied the liquid-gas transition in a system which can be considered as the colloidal analogue of the ionic fluid, i.e. a 1:1 mixture of oppositely charged colloids. The coexistence region is found in the low density--low temperature region, with 3D-Ising criticality. The critical temperature shows a non monotonic behavior with the range of the interaction, $\kappa$; increasing in low values of $\kappa$  and decreasing for the higher ones. This prediction implies a closed region of liquid-gas demixing at constant temperature in the $\kappa - \rho$ plane. The condensed phases are arranged in such a way that each particle is surrounded by shells of particles with alternating charge. The internal energy shows that the overall trend of the critical temperature is lead by the energy gain in forming a dense phase. Liquid-gas separation at long ranges is hindered by the repulsion between similarly charged particles bonded to a third one (of opposite sign), whereas at high $\kappa$ this repulsion is negligible.

\acknowledgments

J.B.C, A.M.P., A.F.B. and F.J.N. acknowledge the financial support by MCyT, under projects no. MAT 2003-03051-CO3-01 and MAT 2004-03581, and J.M.R.E. and L.F.R. under projects no. PB97-0712 and BQU2001-3615-CO2-02.


\begin{thebibliography}{99}
\bibitem{note3} $B_2$ and $B_3$ are calculated by MonteCarlo integration, from their definitions. The full interaction potential is included in the calculation, and the hard sphere contributions are substracted (included in $P_{HS}^{CS}(\rho)$).
\end{thebibliography}
\end{document}